\documentclass{article}
\usepackage{spconf,amsmath,graphicx}
\makeatletter
\usepackage{enumitem}
\setlist{nosep, leftmargin=14pt}

\usepackage{mwe} % to get dummy images
\usepackage{booktabs}

\title{Visual attention analysis of pathologists examining whole slide images of Prostate cancer}

\def\@name{ \emph{Souradeep Chakraborty$^{1}$},  \emph{Ke Ma$^{1}$}, \emph{Rajarsi Gupta$^{2}$}, \emph{Beatrice Knudsen$^{3}$}, \emph{Gregory  J. Zelinsky$^{1,4}$}, \\ \emph{Joel H.  Saltz$^{2}$}, \emph{Dimitris Samaras$^{1}$}}
\vspace{2mm}
\address{$^{1}$ Department of Computer Science, Stony Brook University, NY, USA\\
$^{2}$ Department of Biomedical Informatics, Stony Brook University,  NY, USA \\$^{3}$ Department of Pathology, University of Utah School of Medicine, Utah, USA\\
$^{4}$ Department of Psychology, Stony Brook University,  NY, USA}

\begin{document}

\maketitle

\begin{abstract}
We study the attention of pathologists as they examine whole-slide images (WSIs) of prostate cancer tissue using a digital microscope. To the best of our knowledge, our study is the first to report in detail how pathologists navigate WSIs of prostate cancer as they accumulate information for their diagnoses. We collected slide navigation data (i.e., viewport location, magnification level, and time) from 13 pathologists in 2 groups (5 genitourinary (GU) specialists and 8 general pathologists) and generated visual attention heatmaps and scanpaths. Each pathologist examined five WSIs from the TCGA PRAD dataset, which were selected by a GU pathology specialist. We examined and analyzed the distributions of visual attention for each group of pathologists after each WSI was examined. To quantify the relationship between a pathologist’s attention and evidence for cancer in the WSI, we obtained tumor annotations from a genitourinary specialist. We used these annotations to compute the overlap between the distribution of visual attention and annotated tumor region to identify strong correlations. Motivated by this analysis, we trained a deep learning model to predict visual attention on unseen WSIs. We find that the attention heatmaps predicted by our model correlate quite well with the ground truth attention heatmap and tumor annotations on a test set of 17 WSIs by using various spatial and temporal evaluation metrics. 
\end{abstract}
% %
\section{Introduction}
\label{sec:intro}

Research on attention tracking in digital histopathology images has been evolving \cite{brunye2020eye,sudin2021eye,brunye2017accuracy} in the field of medical imaging. Being able to analyze and predict the visual attention behavior of a pathologist during the examination of WSIs is useful in developing computer-assisted training and clinical decision support systems. For example, pathologists in training  can benefit from visualizing and comparing their attention behavior to experienced pathologists with specialty expertise with the goal of improving interobserver variability in tasks such as Gleason grading in prostate cancer. 

Earliest works to interpret  attention behavior of pathologists as they view WSIs of cancer tissue samples,  include \cite{bombari2012thinking} that conducted eye tracking studies to determine the effect of tumor architecture on the grading of prostate cancer images and \cite{raghunath2012mouse} that examined the spatial coupling between eye-gaze and mouse cursor positions during WSI viewing, and showed that mouse movement tracking data could be a  reliable indicator of a physician's attention and diagnostic behavior. \cite{brunye2017accuracy} explored the complex interactions between pathologists and WSIs that guide diagnostic decision-making using eye-tracking studies. The study in \cite{mercan2018characterizing} used a web-based viewer to record pathologist behavior in order to characterize diagnostic search patterns (scanning and drilling) using viewport attention data from a digital microscope. Similarly, we also used the viewport location, time stamp, and zoom level data in our study. More recently, the works of \cite{brunye2020eye} and \cite{sudin2021eye} have highlighted the importance of  eye tracking and revealed  expertise-related differences in medical image inspection and diagnosis. %While these works have presented behavioral analysis of attention allocation, 
However, significantly less focus has been allocated on visualizing the %spatial and 
spatiotemporal distribution of visual attention in relation with the tumor regions. Also, previous studies have not presented a model that can predict visual attention of pathologists during whole slide image viewing. 

\begin{figure*}
\centering
\includegraphics[width = 13.25cm]{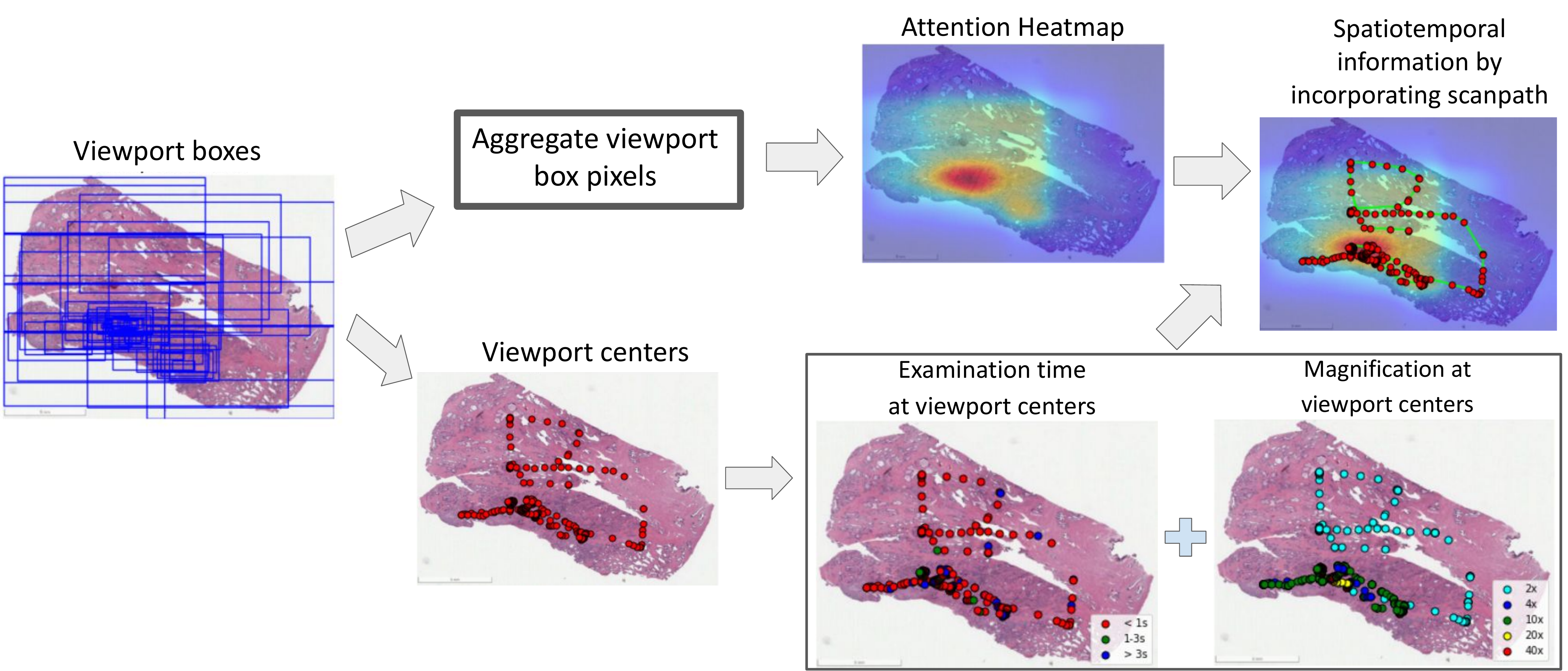}
\vspace{-3mm}
\caption {Processing of the captured attention data to obtain visual attention heatmaps and scanpaths.}
\label{fig:data_processing}
\vspace{-3mm}
\end{figure*}

In this paper, we provide a detailed analysis of how WSIs of prostate cancer were examined by pathologists, which is further correlated with tumor annotations to analyze the differences in the viewing patterns of GU specialists and general pathologists. To the best of our knowledge, we are the first to study the visual attention of pathologists and %present a model that 
predict the distribution of attention during the histopathologic evaluation of prostate cancer in WSIs. %In this study, 
We measured viewing behavior by collecting slide navigation data  while pathologists examined  prostate cancer WSIs as a reasonable surrogate to capture visual attention. We constructed visual attention heatmaps from capturing viewports during examination that we then compared to tumor regions annotated by a GU expert in order to ascertain where the pathologists focused their attention. We also analyzed how visual attention varies across different magnification levels. We used various evaluation metrics  to compare the spatial and spatiotemporal attention distribution  to the annotated tumor regions. Our study %demonstrates the value and 
is proof of concept for collecting slide navigation data to generate surrogate visual trajectories and associated attention heatmaps to understand how pathologists identify cancer presence  in WSIs. Future studies will %further examine %the nuances of 
%viewing behaviors to 
explore intra- and inter-observer variability during the evaluation of more complex cancer features. 
\vspace{-2.2mm}

\section{Methods}
\label{sec:method}
%Here we explain our attention data capture and processing pipeline for whole slide viewing of Prostate cancer tissues. 
\vspace{-3mm}
\subsection{Data collection}
We used QuIP caMicroscope, a web-based toolset for digital pathology data management and visualization \cite{saltz2017containerized} to record the attention data of 13 pathologists as they viewed Prostate cancer whole slide images. % 13 pathologists  participated in our attention tracking study. 
We asked each %pathologist 
participant to provide their expertise level as: (1) boarded and practising anatomic pathologist, (2) resident or fellow. Our  data came from 5 Genitourinary (GU) specialists and 8 general pathologists. Each  subject was instructed to view %a series of 
five whole side images on our web based viewing interface. A GU specialist selected the five WSIs we used for our study   among 342 WSIs of the TCGA-PRAD dataset \cite{zuley2016radiology}. Average viewing time  per slide per pathologist was 103 seconds. No compensation was provided to the pathologists for participation.

Also, in order to compare our attention  data with tumor evidence, we collected tumor region annotations from a GU specialist on the five WSIs in this study. Please see these tumor annotations in Sec.~\ref{sec:results} (second column in  Fig.~\ref{fig:teaser}(a)).

\vspace{-3mm}
\subsection{Data processing}
Our attention data provides us information about where and how long pathologists looked at, and the gaze shifts they made from one region to another while viewing the WSIs. Fig.~\ref{fig:data_processing},  shows our pipeline to process the collected attention data (using viewport boxes). Our attention data comes in two forms: (1) attention heatmaps, (2) attention scanpaths.

\noindent \textbf{Attention Heatmap}: The attention heatmap shows the aggregate spatial distribution of pathologist attention %allocated by pathologists 
during WSI viewing. We assign a value of 1 at every image pixel within a viewport box and sum up the values over all viewport boxes to construct our attention heatmap. Next, we normalize this heatmap to obtain the final attention heatmap, which is: 
% \vspace{-1.5mm}
\begin{equation}
\begin{aligned}
    HM_{Attn.}'^I(x,y) = G^{\sigma}* \sum_{v=1}^V (\sum_{v_x^s}^{v_x^e}\sum_{v_y^s}^{v_y^e}1) \\
    HM_{Attn.}^I =  \frac{HM_{Attn.}'^I - min(HM_{Attn.}'^I)}{max(HM_{Attn.}'^I)-min(HM_{Attn.}'^I)}
\end{aligned}
\label{eq:1}
% \vspace{-2.5mm}
\end{equation}

\noindent where, $HM'^I_{Attn.}$ is the intermediate attention heatmap, $HM_{Attn.}^I$ is the final normalized attention heatmap, $V$ is the number of viewport boxes on a WSI $I$, and $v_x^s$, $v_x^e$,$v_y^s$,$v_y^e$ denote the starting and the ending $x$ and $y$ coordinates of the viewport box $v$ respectively. $G^{\sigma}$ is a 2D gaussian  with $\sigma = 16$ pixels that smooths the intermediate heatmap $HM'^I_{Attn.}$.\\

\noindent \textbf{Attention Scanpath}: The attention heatmap only captures  aggregate spatial attention distribution. %, it does not provide any information about 
%but not the temporal order in which various slide regions are viewed. %visited during viewing. 
In order to capture the viewing trajectories, we produce attention scanpaths from the collected viewport data. We stack the viewport centers of every viewport box, $v$ in WSI $I$ in order to construct our scanpath, $SP_{Attention}^I$ as $SP_{Attention}^I = [v_{center}^1,v_{center}^2,...,v_{center}^{(V-1)},v_{center}^V]$, where the viewport center of a viewport box $i$ is
$v_{center}^i = (\frac{v_x^s+v_x^e}{2},\frac{v_y^s+v_y^e}{2})$. 

In Fig.~\ref{fig:data_processing}, we also show the viewing examination time and the magnification levels at the viewport centers. In the depicted WSI instance,  the pathologist spent less than 1 second at most of the viewports (indicated by the red viewport centers) and mostly  viewed the slide at 2X and 4X magnification. 
% \vspace{-4mm}
\subsection{Predicting attention heatmaps using deep learning}

Here we discuss the deep learning model we trained for predicting attention heatmap over a WSI. We formulate attention prediction as a classification task where the goal is to classify a WSI patch into one of the $N$ attention intensity bins ($N=5$ in our study). During training, we discretize the average pixel intensity of every heatmap patch into an intensity bin. At inference time, we assign the average pixel intensity of a bin to the image patch according to the predicted class in order to construct the patch-wise heatmaps. We reconstruct the final attention heatmap over the WSI by assembling the predicted patch-wise heatmaps followed by gaussian smoothing and map normalization.

\begin{figure}
\centering
\includegraphics[width = 8.50cm]{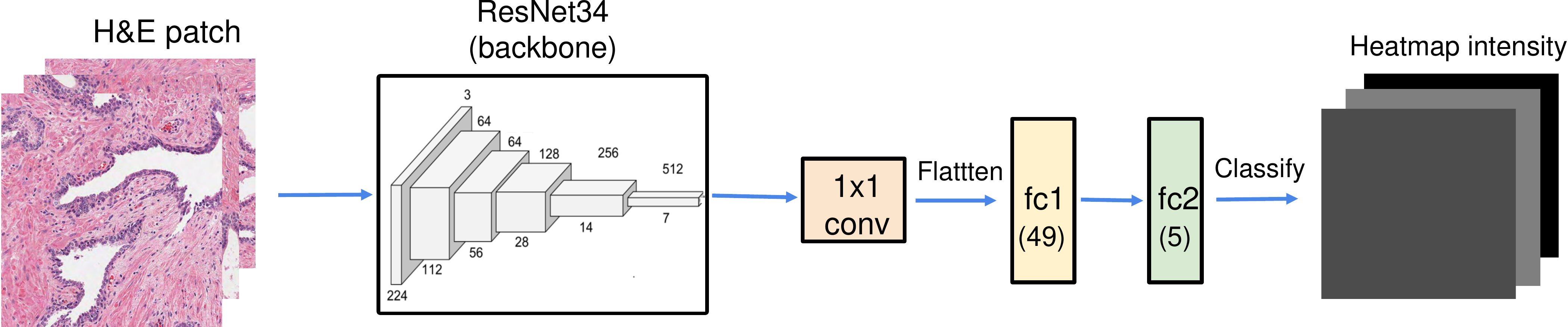}
\caption {Our model, ProstAttNet for predicting visual attention on whole slide images of Prostate cancer.}
\label{fig:network}
\vspace{-5mm}
\end{figure}

We trained a CNN model, ProstAttNet (Prostate AttentionNet), to classify an image patch into one of the 5 attention intensity levels using the pre-trained ResNet34 model \cite{he2016deep} as the backbone followed by a $1 \times 1$ convolutional  layer and two fully connected (fc) layers. We depict our attention prediction model, ProstAttNet in Fig.~\ref{fig:network}. See suppl. for training details.   
\vspace{-2.5mm}
\section{Results}
\label{sec:results}
%We qualitatively and quantitatively compare  attention heatmaps and  scanpaths of the participating pathologists with the ground truth tumor probability map (by a GU specialist).

% \begin{figure*}
% \centering
% \includegraphics[scale = 0.20]{./figures/main_results.pdf}
% \caption {(a) Comparison of the visual attention heatmaps with the ground truth tumor segmentation maps. We see a strong correlation between the two maps. (b) Comparison of the predicted attention heatmap using our ProstAttNet model with the ground truth segmentation map and the ground truth attention heatmap constructed from attention data collected from a Genitourinary specialist on 4 test WSI instances.}
% \label{fig:teaser}
% \end{figure*}
% % width = 7.00cm

\subsection{Qualitative Evaluation}
\begin{figure}[h]
\centering
\includegraphics[scale = 0.17]{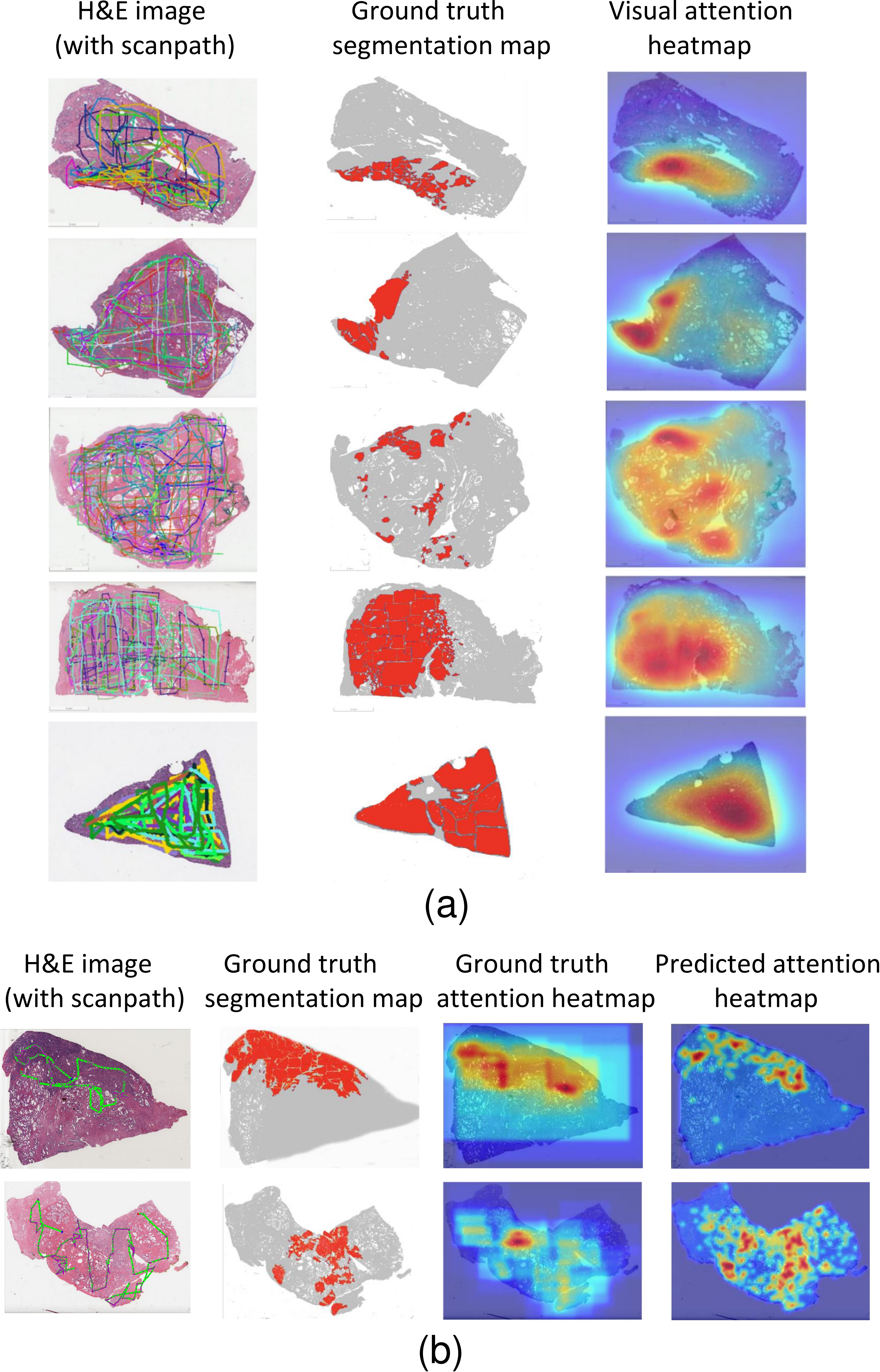}
\caption {(a) Comparison of the visual attention heatmaps with the ground truth tumor segmentation maps. We see a strong correlation between the two maps. (b) Comparison of the predicted attention heatmap using our ProstAttNet model with the ground truth segmentation map and the ground truth attention heatmap constructed from attention data collected from a Genitourinary specialist on two test WSI instances.}
\label{fig:teaser}
\vspace{-2mm}
\end{figure}

Fig.~\ref{fig:teaser}(a) shows  ground truth tumor segmentation maps and  attention heatmaps for the five H\&E WSIs  of this study. Attention heatmaps have a high spatial correlation with tumor locations in the ground truth tumor segmentation map. We also predicted attention heatmaps using our ProstAttNet model on a test dataset of 17 whole slide images (from the TCGA-PRAD dataset). Fig.~\ref{fig:teaser}(b) compares prediction results on two WSI instances from our test dataset. The ground truth attention heatmap was constructed from slide navigation data collected from th GU specialist only for the purpose of validating model predictions. We see that the predicted attention heatmap correlates well with the ground truth attention heatmap and the ground truth tumor segmentations.
% Please see  suppl. material for  detailed visualizations of attention heatmaps, scanpaths and average viewing times of the two pathologist groups.

\begin{figure}[h]
\centering
\includegraphics[scale=0.29]{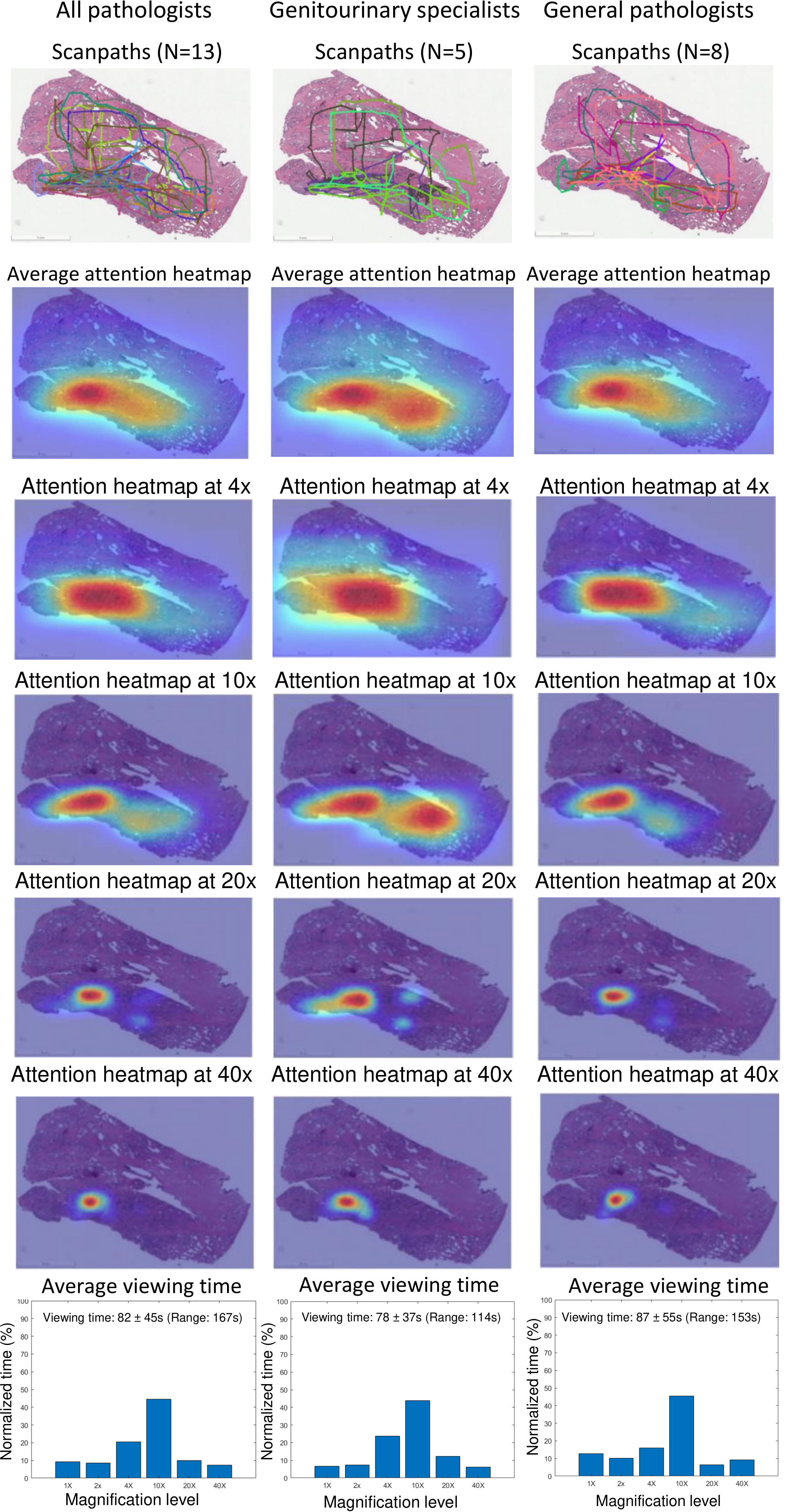}
\caption {Comparison of attention heatmaps (average and magnification level-wise), scanpaths and average viewing time of the Genitourinary specialists and the general pathologists on the TCGA-EJ-7328 WSI from the TCGA-PRAD dataset.}
\label{fig:gen_gu_comp}
\vspace{-3.5mm}
\end{figure}

We also compared the viewing behavior of the GU specialists and the general pathologists in our study. In Fig.~\ref{fig:gen_gu_comp}, we show how pathologists examined WSIs of Prostate cancer by overlaying navigation scanpaths on H\&E images and generating  attention heatmaps to compare overall behavior of all pathologists, which is subdivided to evaluate potential differences in behavior between genitourinary (GU) specialists and general pathologists for one of the cases in our study. We also generate attention heatmaps with respect to the viewport magnification level to evaluate visual behavior at 4X, 10X, 20X, and 40X. We see a high concurrence in the viewed regions at every magnification level among both groups of pathologists. In addition, we also show the average viewing time of the two groups of pathologists across four different magnification levels. While we see good concurrence within the groups in terms of the attended regions, we also observe some differences in their averaged attention heatmaps. While the general pathologists did not seem to have allocated sufficient attention to the bottom right tumor regions, the GU specialists looked at more of the tumor region (left and right high intensity areas) as seen in the overall averaged heatmap in the second row.

% \begin{figure}
% \centering
% \includegraphics[width = 10.20cm]{./figures/test_pred.pdf}
% \centering
% \caption {Attention heatmaps predicted by our Pr-AttentionNet model on three instances from our test dataset. The yellow contours indicate the tumor regions indicated by a pathologist.}
% \label{fig:test_pred}
% \end{figure}

% \begin{figure}
% \centering
% \includegraphics[width = 9.10cm]{./figures/fig2.pdf}
% \caption {Attention heatmap by different magnification levels on the TCGA-EJ-7328 image}
% \label{fig:mag_levels}
% \end{figure} 
\vspace{-3mm}
\subsection{Quantitative Evaluation}

To evaluate how well attention data  predicts tumor regions (from pathologists' annotations), we used the Cross correlation (CC) metric to compare our attention heatmaps and the Semantic Sequence Score (SSS) metric to evaluate the scanpaths constructed from the viewport centers. We compared the attention heatmaps with tumor proabability maps constructed from the tumor annotations. All compared maps were downsampled by a factor of 1/16  (compared to original image size) for computational reasons. We convolved the tumor segmentation map with a 2D gaussian  ($\sigma$ = 16 pixels) and normalized this map to obtain the tumor probability map for comparison with the attention heatmap. In order to ensure that the distributions of the attention heatmap and the tumor probability map are similar, we perform histogram matching \cite{gonzales1977gray} of the two maps as a pre-processing step  \cite{peacock2020center}. 

We use the cross-correlation score (CC) by interpreting the attention heatmap,  $HM_{Atten.}$ and the tumor probability map, $PM_{Tumor}$ as
random variables and measure the linear relationship
between them. High positive CC values
occur  where the compared maps  have values of similar magnitude. The Semantic Sequence Score (SSS) metric we use captures the inter-observer scanpath similarity across the two groups of pathologists in terms of the grade of tumor regions traversed during whole slide image viewing. To obtain SSS we modified the Sequence Score (SS) metric \cite{borji2013analysis} (generally used to compare scanpaths on natural images) by substituting clusters based on eye fixations with clusters based on Gleason grades. Specifically, we first convert a scanpath into a string representing the sequence of Gleason grades corresponding to the viewport box centers (e.g. $G_3$-$G_5$-$G_4$-$G_3$, $G_4$-$G_4$-$G_3$-$G_5$-$G_5$, etc. where $G_n$ represents Gleason grade $n$). We then use a string matching algorithm \cite{needleman1970general} 
to measure string similarity.

\begin{table}[]
\centering
\resizebox{\linewidth}{!}{%
\begin{tabular}{@{}lcccccc@{}}
\toprule
Case & \multicolumn{2}{c}{All pathologists} & \multicolumn{2}{c}{GU specialists} & \multicolumn{2}{c}{General pathologists} \\
\cmidrule(lr){2-3}
\cmidrule(lr){4-5}
\cmidrule(lr){6-7} 
             & CC    & SSS   & CC    & SSS   & CC    & SSS   \\ \midrule
TCGA-EJ-7328 & 0.729 & 0.421 & 0.765 & 0.390 & 0.706 & 0.441 \\
TCGA-HC-A8D1 & 0.877 & 0.412 & 0.881 & 0.464 & 0.874 & 0.380 \\
TCGA-G9-6384 & 0.712 & 0.481 & 0.725 & 0.562 & 0.704 & 0.430 \\
TCGA-EJ-7315 & 0.780 & 0.390 & 0.787 & 0.445 & 0.776 & 0.355 \\
TCGA-G9-6494 & 0.473 & 0.398 & 0.437 & 0.496 & 0.495 & 0.336 \\
\midrule
Average      & 0.714 & 0.420 & 0.719 & 0.472 & 0.711 & 0.388 \\ \bottomrule
\end{tabular}%
}
\caption{Quantitative evaluation of attention data and tumor annotations (by pathologist),
CC is the Cross-Correlation score for comparing attention heatmap with ground truth segmentation map and SSS is Semantic Sequence Score comparing inter-observer similarity on attention scanpaths.}
\label{tab:table1}
\vspace{-4mm}
\end{table}

% \begin{figure}
% \centering
% \includegraphics[width = 8.20cm]{./figures/metrics.pdf}
% \caption {Quantitative evaluation of attention data and tumor annotations (by pathologist), %using different spatial and spatio-temporal metrics. Here
% CC is the Cross-Correlation score for comparing attention heatmap with ground truth segmentation map and SSS is the Semantic Sequence Score metric comparing inter-observer similarity on attention scanpaths.}
% \label{fig:table}
% \end{figure}

Table.~\ref{tab:table1} lists the different evaluation metrics discussed above for the GU specialists and the general pathologists on the five WSIs. A high average cross correlation score of 0.714 across all pathologists suggests a strong spatial correlation between the distribution of attention and the ground truth tumor locations. Despite slight differences in the viewing behaviors of the two groups, independent t-test analysis indicates that  differences in the degree of correlation between tumor region and attention heatmaps are not significant ($p > 0.52$ averaged over the 5 images). However, the consistently high SSS for  GU specialists compared to  general pathologists indicates more consistent viewing behavior within this group compared to  general pathologists. Using  ProstAttNet, we obtained average CC score of 0.453  between  predicted attention and  ground truth attention heatmaps on our test set of 17 WSIs. The predicted attention heatmap overlapped with the ground truth segmentation map with a CC score of 0.532.

\vspace{-2.5mm}
\section{Conclusion}
\label{sec:conclusion}
We have shown how pathologists allocate attention while viewing   prostate cancer WSIs and presented a deep learning model that predicts visual attention. Our data visualization schema allow us to understand how pathologists examine WSIs to identify cancer. We find a strong correlation between the locations of tumor, annotated by a GU specialist, and attention heatmaps of 13 pathologists. In the  future, we will collect attention data in a larger study with more WSIs in order to improve our attention prediction model. Also, we plan to leverage attention data to train a deep learning model for tumor segmentation and Gleason grading to test whether we can circumvent the need for extensive annotation.
% \vspace{-2mm}
\section{COMPLIANCE WITH ETHICAL STANDARDS}
Data collection was conducted within the ethical guidelines established and overseen by the Stony Brook University Institutional Review Board.%Data collection was conducted within the ethical guidelines established by the Institutional Review Board responsible for overseeing human subjects research at Stony Brook University.
 
% Gaze collection was made using sbu irb..
% (Ask Greg and Ke for acknowledging data collection, ask Beatrice if she had any grants or people who helped her)
% \vspace{-3mm}
\section{ACKNOWLEDGEMENTS}
This work was supported by a seed grant from the Stony Brook University Office of the Vice President for Research (1150956-3-63845), NSF grant IIS-2123920, NCI grant U24CA215109 and generous private  support from Bob Beals and Betsy Barton.

% \bibliographystyle{IEEEbib}
% \bibliography{strings,refs}

\vspace{4mm}

\section{Supplementary: Training details of ProstAttNet}

During training of our ProstAttNet model, we froze the pre-trained ResNet34 weights and allowed gradient flow over the fully connected layers of our model. For training the model, we used image patches (each of size $500 \times 500$) extracted from four whole slide images at 10x magnification (the most frequent magnification level used by pathologists during WSI  viewing per our analysis). The remaining single slide was used for validation. We generated a total of 58.5K image patches for training with labels corresponding to 13 pathologists per patch. We also performed data augmentation %following 
\cite{tellez2018whole} by introducing color jitter, random horizontal and vertical image flips during training. %, which helped us train a well generalized model. 
We trained this model using the Cross-Entropy loss between the predicted and the ground truth heatmap intensity bin. We used the Adam optimizer \cite{kingma2014adam} with  initial learning rate of 0.005. Training converged within 20 epochs with a total training time of 8.5 hours on a Nvidia Titan-Xp GPU. We used the PyTorch deep learning library for the model implementation.

\bibliographystyle{IEEEbib}
\bibliography{strings,refs}

\end{document}